\documentclass[twocolumn,showpacs,preprintnumbers,amsmath,amssymb]{revtex4}
\usepackage{graphicx,textcomp}% Include figure files
\usepackage{dcolumn}% Align table columns on decimal point
\usepackage{bm}% bold math

\begin{document}

\preprint{APS/123-QED}

\title{Cold Reactions of Alkali and Water Clusters inside Helium Nanodroplets}% Force line breaks with \\

\author{S. M\"uller$^1$}
%\author{C.P. Schulz$^2$}
\author{S. Krapf$^2$}
\author{Th. Koslowski$^2$}
\author{M. Mudrich$^1$}
%\email{Marcel.Mudrich@physik.uni-freiburg.de}
\author{F. Stienkemeier$^1$}

\affiliation{$^1$Physikalisches Institut, Universit\"at Freiburg, 79104 Freiburg, Germany}
%\affiliation{$^3$Max-Born-Institut, Berlin, Germany}
\affiliation{$^2$Institut f\"ur Physikalische Chemie, Universit\"at Freiburg, 79104 Freiburg, Germany}

\date{\today}% It is always \today, today,

\begin{abstract}
\noindent
The reaction of alkali (Na, Cs) clusters with water clusters embedded in helium nanodroplets is studied using femtosecond photo-ionization as well as electron impact ionization. Unlike Na clusters, Cs clusters are found to completely react with water in spite of the ultracold helium droplet environment. Mass spectra of the Cs$_n$+(H$_2$O)$_m$ reaction products are interpreted in terms of stability with respect to fragmentation using high-level molecular structure calculations.

\end{abstract}


\maketitle

Chemical reactivity at very low temperatures is a topic of increasing interest both for theory and for experiment~\cite{Krems:2005,HudsonPRA:2006,Toennies:2007,Willitsch:2008}. On the one hand, this is due to new phenomena which arise as the reaction dynamics enters the quantum scattering regime, \textit{e.\,g.} quantum scattering resonances in the reaction cross sections~\cite{Krems:2005}.
On the other hand, the recent progress in developing experimental techniques for producing and trapping cold and ultracold molecules allows for extending experiments on reactive scattering to the quantum regime. Cold reactive ion-molecule collisions have recently been reported~\cite{Willitsch:2008} and experiments aiming at studying cold reactive neutral molecule-molecule collisions using Stark-decelerated molecular beams are in preparation~\cite{Gilijamse:2006,HudsonPRA:2006}.

Helium (He) nanodroplets are particularly well suited as a nearly ideal cryogenic matrix for studying reactions at sub-Kelvin temperatures~\cite{Toennies:2004,Toennies:2007}. The high mobility of dopant atoms and molecules inside the superfluid droplets allows to bring together atoms or molecules of rather different nature,
%\textit{e.\,g.} high refractory metal atoms and fragile organic molecules
the inert He environment causing only negligible perturbations~\cite{Mudrich:2007,Toennies:2004}. Fast dissipation of internal excitations by evaporation of He atoms leads to efficient cooling of the internal degrees of freedom and allows to prepare weakly-bound precursor states as starting conditions prior to initiating a chemical reaction, \textit{e.\,g.} by shining in a laser pulse. Thus, using femtosecond (fs) pump-probe spectroscopy, the formation dynamics of alkali metal (M)-He exciplexes on He nanodroplets has been studied in real time~\cite{Mudrich:2008,Droppelmann:2004}. The dynamics of photo-induced dissociation inside He nanodroplets has recently been investigated using ion imaging techniques~\cite{Braun:2004}.
%hotoinduced chemical dynamics of high-spin alkali trimers~\cite{Higgins:1996}.

So far, the only chemical reaction involving neutral heterogeneous reactant molecules inside He nanodroplets is Ba + N$_2$O $\rightarrow$ BaO + N$_2$, which was characterized by chemiluminescence spectroscopy~\cite{Lugovoj:2000}.
Recently, some efforts for exploring chemical reactions in He droplets have been made \textit{e.\,g.} on the reaction F + CH$_4$ \cite{Merritt:2006}. However, no reactive process was found indicating that at the low temperature of the He droplet environment even small energy barriers may impede chemical reactions. Beside the reaction of neutrals, a rich ion-molecule chemistry inside ionized He droplets has been reported~\cite{Toennies:2004}. \textit{E.\,g.}, secondary reactions of N$_2^+$, D$_2^+$ and CH$_4^+$ fragment ions created by electron impact ionization (EII) of the He have been identified, thus evoking the term 'flying nano-cryo-reactors'~\cite{Farnik:2005}.

The showcase reaction Na+H$_2$O has been studied in great detail at high temperatures to solve the long-standing puzzle~\cite{Hertel:1991,Bewig:1998,Buck:1998,Mundy:2000,Bobbert:2001,Steinbach:2005}: While M metals are well-known to react with water in a violent exothermic reaction at the macroscopic scale, single Na atoms colliding with water molecules or even with clusters do not react besides being solvated. Only when combining Na$_n$ clusters with (H$_2$O)$_m$ clusters, chemical reactions were found to take place and products Na(NaOH)$_n^+$ appeared in the product mass spectra~\cite{Bobbert:2001,Steinbach:2005}. Both experimental and theoretical results point at a complex elementary reaction involving three Na atoms and six H$_2$O molecules~\cite{Mundy:2000,Steinbach:2005}.
In this Letter we report on the chemical reaction of alkali (M=Na, Cs) clusters with water (H$_2$O) clusters inside He nanodroplets. We have done similar studies using K and Rb; these results nicely go along with the interpretations given below and will be published elsewhere.

\begin{figure}[t!]
\begin{center}
{
\includegraphics[width=0.51\textwidth]{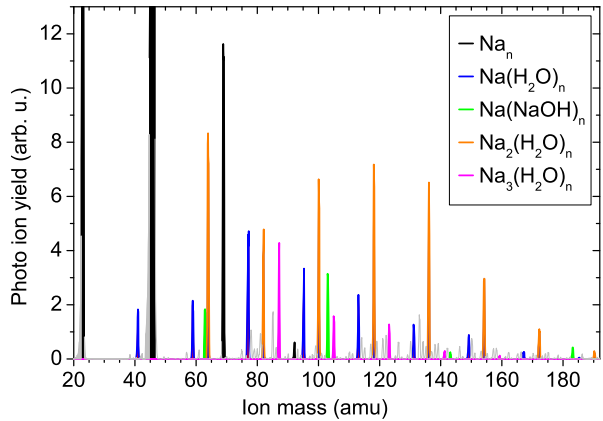}}
\caption{Photo ionization mass spectrum of mixed sodium (Na)-water (H$_2$O) clusters in He nanodroplets. The dominant cluster series Na(H$_2$O)$_n$, Na(NaOH)$_n$ and Na$_2$(H$_2$O)$_n$ are highlighted.
%Some peaks are due to alkali impurities in the sample, e.\,g. KNa$_\text{2}$ or Cs.
}
\label{NaWater}
\end{center}
\end{figure}

Our experiment combines a beam of He nanodroplets, consecutively doped with water molecules and with M atoms, with fs laser ionization or with EII. The experimental setup is described in detail elsewhere~\cite{Mudrich:2008,Droppelmann:2004}. In short, ultrapure He gas is expanded at high pressure (60\,bar) out of a cryogenic nozzle (diameter 10\,$\mu$m) into high vacuum. At the nozzle temperature of 13\,K the average droplet size amounts to about 20,000 atoms. In the doping chamber the beam passes through two scattering cells for the droplets to pick up water molecules and subsequently M atoms. For reaction products to appear in the photo ionization (PI) mass spectrum, the vapor pressures in the pick-up cells have to be increased significantly above the levels needed for doping one M atom or H$_2$O molecule. At a vapor pressure corresponding to maximum probability for picking up 10 M atoms per droplet, cluster sizes up to M$_6$ are detected without adding H$_2$O mainly due to fragmentation induced by the ionization process. The H$_2$O doping conditions are chosen such that a broad (FWHM $\approx \bar{N}$) cluster size distribution peaked around the prominent mass peak at $N$=20 water molecules per cluster is observed in the EII mass spectrum without doping with M atoms.
%he water 20-mer is well-known for its extraordinary stability due to its highly regular geometrical structure~\cite{Wu:2005}.

In order to obtain a realistic picture of the compounds produced in the reaction, two complementary ionization schemes are employed, non-resonant multi-photon PI using laser pulses of 150 fs duration at 860 nm (755 nm) for the Cs (Na) experiments and EII at 30 eV electron energy. While fs photo ionization acts on the M atoms and clusters due to their low ionization potentials ($\sim$2.7-5 eV), it is completely inefficient for pure water clusters. In contrast, EII proceeds in a two-step process mediated by the He atoms. Thus, (H$_2$O)$_m$ clusters, which are immersed into the droplets, are easier to ionize by EII than the neat clusters M$_n$, which stay bound to the droplet surface.

The resulting PI mass spectrum of He droplets simultaneously doped with Na$_n$ and (H$_2$O)$_m$ is displayed in Fig.~\ref{NaWater}. It is clearly dominated by small clusters Na$_{1-3}$ carrying up to 8 water molecules reflecting weakly bound van der Waals-complexes (blue and red lines)~\cite{Douberly:2007}. The only compounds originating from the reaction with H$_2$O are low-abundant Na hydroxide masses, Na(NaOH)$_n$, the most prominent peak being the one of Na(NaOH)$_2$ (green line). This finding is in line with cluster reactions being largely frozen out in the ultracold droplet environment, except for a small fraction that has not completely thermalized to 0.4\,K before initiating the reaction.

 \begin{figure}[t!]
\begin{center}
{
\includegraphics[width=0.51\textwidth]{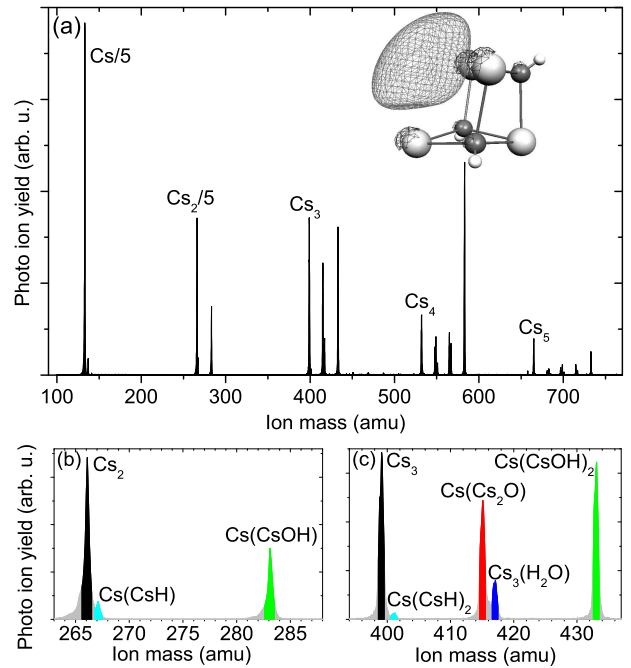}}
\caption{a) Photo ionization mass spectrum of mixed cesium-water clusters in He nanodroplets. The vertical scale is shrunk by factor 5 in the mass range 100 to 300\,amu for the sake of clarity. The inset illustrates the molecular structure of the prominent compound Cs(CsOH)$_3^+$ (see text). b) and c) Detailed views of the mass spectrum showing Cs$_2$ and Cs$_3$, respectively, and associated reaction products.
}
\label{CsWater_MS}
\end{center}
\end{figure}

However, a totally different picture presents itself when Cs$_n$ clusters are combined with H$_2$O clusters inside He droplets. The PI mass spectrum, shown in Fig.~\ref{CsWater_MS}, features a number of additional peaks aside from the pure Cs$_n$ masses, which follow a strikingly regular pattern: Roughly speaking, every Cs$_m$ cluster mass peak is augmented by $m-1$ satellites with additional masses of about 1 to $m-1$ water molecules. Except for the Cs$_2$-satellite, the last peak of each series, corresponding to Cs(CsOH)$_m$, shows extraordinarily high abundance, in the case of Cs(CsOH)$_3$ even exceeding the one of the corresponding neat cluster Cs$_4$. However, no van der Waals-bound complexes, Cs$_n$(H$_2$O)$_m$ with $m\ge n$ are observed under any conditions, in contrast to the Na+H$_2$O system.

Upon closer inspection the satellite peaks reveal a more complex composition reflecting different reaction products, shown in Fig.~\ref{CsWater_MS} b) and c). Besides the dominating Cs(CsOH)$_m$-peaks we find various Cs oxide, hydride, and hydroxide compound clusters as well as reaction products involving H$_2$O complexes, which have not been observed in earlier experiments. These compounds are found to follow odd-even alternations in terms of the number of Cs atoms involved. \textit{E.\,g.}, Cs$_n$O and Cs$_n$H$_2$O are most abundant for odd $n$, whereas Cs$_n$OH mass peaks are more pronounced for even $n$. This suggests the notation used in legends of Fig.~\ref{CsWater_MS} a)-c), in accordance with valencies of the atomic and molecular constituents. These compounds may reflect intermediate states of the reaction which are stabilized by the dissipative He droplet environment. Note that NaH formation was identified as the rate determining step with a reaction barrier of 0.6\,eV in the Na$_n$+(H$_2$O)$_m$ reaction~\cite{Mundy:2000}.

The surprising result of this experiment is that small clusters Cs$_n$ actually react very efficiently with water inside He nanodroplets, despite the low temperature environment. This implies nearly vanishing activation energy for cluster reactions involving Cs, in contrast to the findings for the elementary Na reaction~\cite{Mundy:2000}. Possibly, surplus energy from the ionization energy or high-spin states of Cs clusters, as expected from the aggregation process in He droplets, may facilitate reactions. The lack of H$_2$O appendages may also be related to the lower hydration energy of Cs atoms (2.86\,eV per molecule for bulk water) as compared to the one of Na atoms (4.21\,eV). While water molecules are bound rather tightly to Na$_n$, the surplus energy from ionization may boil off the excess water from Cs$_n$ clusters more easily. Strong reactivity of Cs$_n$ with H$_2$O in He droplets is even more unexpected since Cs atoms and molecules are located in surface states whereas water molecules are immersed inside the droplets. Apparently, long-range attractive forces bring the Cs and H$_2$O clusters together, as previously observed with M atoms and HCN molecules~\cite{Douberly:2007}.

\begin{figure}[t!]
\begin{center}
{
\includegraphics[width=0.51\textwidth]{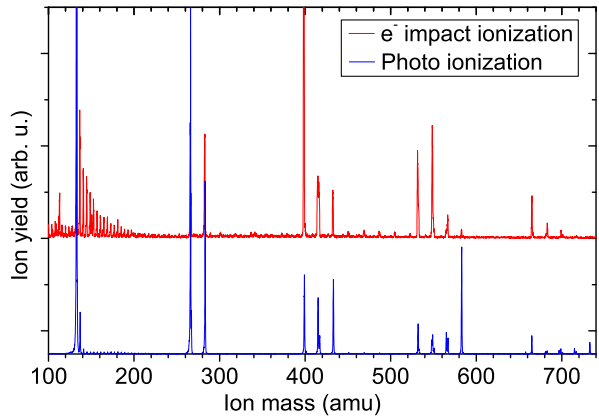}}
\caption{Mass spectra of mixed cesium-water clusters in He nanodroplets, recorded using electron impact ionization (upper trace) vs. fs photo ionization at 860\,nm (lower trace). Despite the different ionization mechanisms (see text) the spectra show great qualitative resemblance.
}
\label{CsWater_EII_PI}
\end{center}
\end{figure}

In order to investigate to what extent the observed mass spectra depend on the particular detection scheme, EII is applied under identical conditions as the PI measurements. Fig.~\ref{CsWater_EII_PI} illustrates the remarkable qualitative agreement of the mass spectra obtained for the two complementary ionization schemes. In both cases the mass spectra consist of progressions of neat clusters Cs$_m$ with $m-1$ 'water' satellites. Thus, the interpretation of the excess Cs atom in terms of serving as a chromophore which enhances the PI cross section of the compound can be ruled out~\cite{Steinbach:2005}. The similarity of PI and EII spectra show that these spectra actually reproduce the abundance distributions of ionized compounds and that reactions take place among neutral precursor clusters before ionization.  The differences in the mass spectra can be attributed to varying degrees of fragmentation of the individual compound clusters due to the differing excess energies deposited in the clusters. Furthermore, ionization spectra of the individual components (not shown here) clearly confirm this interpretation.

In order to draw more quantitative conclusions, high-level quantum chemical calculations are performed to determine equilibrium structures, ionization potentials (IPs), and stabilities with respect to
fragmentation for the observed clusters. All calculations are done with the
Gaussian 03 program package~\cite{Gaussian03} using Hybrid-HF-DFT with the
PBE1PBE~\cite{Ernzerhof1999,Adamo1999} functional. On Cs, a
semi-relativistic~\cite{Leininger1996} and a relativistic core
potential~\cite{Lim2005} with the corresponding basis sets are used for the
geometry optimizations and for single point calculations on the optimized geometries,
respectively. On H and on second row atoms the 6-31++G(d,p) basis is applied for
the geometry optimizations and the 6-311++G(3df,3pd) basis is used for the
single point calculations. This method is found to reproduce the experimental
atomization energies of Cs$_2$, Cs$_2$O, CsOH, CsH, CsF and
CsCl with a mean deviation of $\le 0.2$ eV.

The theoretical results show that all detected compound masses feature low vertical IP$<3.3$\,eV and are thus amenable to efficient ionization by 2 or 3 photons from our fs-laser. In contrast, the likely reaction products (CsOH)$_n$, have large vertical IP$>6.4$ eV for $n\leq 5$, thus requiring at least 5 photons for laser ionization. However, the fact that these cluster masses are absent in the EII mass spectra as well demands further explanations. It turns out that while (CsOH)$_n$ clusters require high energies $\Delta E>3$ eV for fragmentation, the ionization products (CsOH)$^+_n$ are quite unstable with respect to the likely fragmentation channel, the loss of OH. For all cluster masses with $n\leq 5$ fragmentation energies are $\Delta E<1$ eV and fall below the excess energy upon 5-photon ionization. We conclude that while (CsOH)$_n$ are most likely produced in the M-H$_2$O reactions, reaction products fragment into the observed cluster masses Cs(CsOH)$^+_n$ upon ionization.

\begin{figure}[t!]
\begin{center}
{
\includegraphics[width=0.52\textwidth]{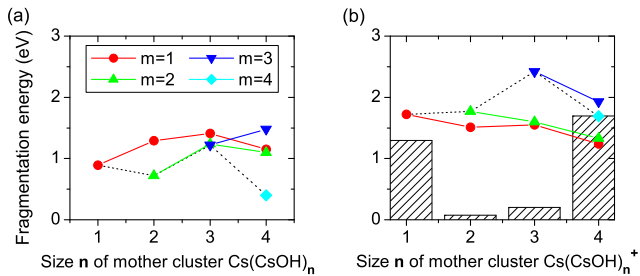}}
\caption{(a) Fragmentation energies for the endothermic fragmentation reaction of neutral clusters Cs(CsOH)$_n\rightarrow$ Cs(CsOH)$_{n-m}$+(CsOH)$_m$ and (b) of cationic clusters Cs(CsOH)$^+_n\rightarrow$ Cs(CsOH)$^+_{n-m}$+(CsOH)$_m$. The bars represent the excess energy after multi-photon ionization of the neutral clusters.}
\label{FragmentEnergies}
\end{center}
\end{figure}

The stability of neutral Cs(CsOH)$_n$ and of cationic Cs(CsOH)$^+_n$ clusters, which dominate the mass spectrum, is theoretically analyzed with respect to the most probable fragmentation channel, the loss of (CsOH)$_m$. Fig.~\ref{FragmentEnergies} depicts the fragmentation energies $\Delta E$. Interestingly, for the neutral clusters the fragmentation channel requiring lowest fragmentation energy is the loss of the entire hydroxide compound, (CsOH)$_{m=n}$, (dashed line in Fig.~\ref{FragmentEnergies} (a)), whereas for the cations this channel is highest in $\Delta E$ (dashed line in Fig.~\ref{FragmentEnergies} (b)), except for Cs(CsOH)$^+_4$. This is due to the large binding energies of neutral (CsOH)$_m$, and to the tendency of the positive charge to spread out over a large cluster to minimize Coulomb repulsion, respectively.

While $\Delta E$ for cationic Cs(CsOH)$^+_n$ is substantially larger than $\Delta E$ for neutral Cs(CsOH)$_n$, surprisingly both the neutrals and the ions show an enhanced stability for $n=3$; $\Delta E$ is particularly high with respect to all 3 possible fragmentation channels. This is due to the regular geometric structure of Cs(CsOH)$^{0/+}_3$, which matches the cuboid rock salt lattice with one unoccupied site, where an electron is localized in the case of neutral Cs(CsOH)$_3$ (see inset in Fig.~\ref{CsWater_MS}). In particular, $\Delta E$ by far exceeds the excess energy deposited in the cluster upon 2-photon ionization, depicted as hatched bars in Fig.~\ref{FragmentEnergies} (b). The same is true for $n=1,\ 2$. The enhanced stability for $n=3$ and the expected fragmentation of Cs(CsOH)$^{+}_4$ into Cs(CsOH)$^{+}_3$ nicely explains the extraordinarily high abundance of that compound observed by laser ionization (Fig.~\ref{CsWater_MS}).

The abundance of cluster masses of the type Cs$_n$H$^+$, Cs$_n$O$^+$, Cs$_n$(H$_2$O)$^+$, and Cs$_n$(OH)$^+$, which show alternating abundance patterns, may be rationalized using the following simple picture. High stability is determined by the number $n$ of valence electrons of the Cs$_n$-part, which either form chemical bonds or pair up in even numbers. \textit{E.\,g.}, Cs$_2$OH$^+$ (Cs$_3$O$^+$) are particularly stable, since one (two) electrons are needed for Cs binding to OH (O), and the remaining electron is ejected upon ionization. This picture, which applies here due to predominantly ionic bonding of these small compound clusters, is confirmed by our theoretical results in terms of cluster stabilities with respect to fragmentation.

In conclusion, we find that while Na and H$_2$O clusters in He droplets predominantly form van der Waals complexes, Cs$_n$ clusters completely chemically react with H$_2$O to form a variety of compounds despite the ultracold droplet environment. The peculiar abundance pattern of reaction products is largely independent of the detection scheme applied. All prominent features are interpreted in terms of stabilities of individual compound clusters with respect to fragmentation, as obtained from high-level molecular structure calculations. In particular, the high abundance of Cs(CsOH)$_3^+$ is found to be related to the extraordinary stability of neutral Cs(CsOH)$_3$ and of its cation as a result of their regular cuboid structure. These experiments demonstrate the potential of He nanodroplets to serve as nanoscopic cryo-reactors for probing elementary reactions at ultralow temperatures.

Stimulating discussions with U. Buck and C.P. Schulz are gratefully acknowledged. This work has been supported by the Deutsche Forschungsgemeinschaft.

\bibliographystyle{apsrev}

%\bibliography{MarcelBib2}
\end{document}